\documentclass{emulateapj}

\begin{document}
\slugcomment{Accepted to ApJ Letters}

\title{ THE ANISOTROPIC SPATIAL DISTRIBUTION OF HYPERVELOCITY STARS }

\author{Warren R.\ Brown,
	Margaret J.\ Geller,
	Scott J.\ Kenyon}

\affil{Smithsonian Astrophysical Observatory, 60 Garden St, Cambridge, MA 02138}
\email{ wbrown@cfa.harvard.edu} %, mgeller@cfa.harvard.edu, skenyon@cfa.harvard.edu}

\and    \author{Benjamin C.\ Bromley}
\affil{Department of Physics, University of Utah, 115 S 1400 E, Rm 201, Salt Lake City, UT 84112}
%\email{bromley@physics.utah.edu}

\shorttitle{ Anisotropic Distribution of Hypervelocity Stars }
\shortauthors{Brown et al.}

\begin{abstract}

	We study the distribution of angular positions and angular separations of
unbound hypervelocity stars (HVSs).  HVSs are spatially anisotropic at the
3-$\sigma$ level.  The spatial anisotropy is significant in Galactic longitude, not
in latitude, and the inclusion of lower velocity, possibly bound HVSs reduces the
significance of the anisotropy.  We discuss how the observed distribution of HVSs
may be linked to their origin.  In the future, measuring the distribution of HVSs in
the southern sky will provide additional constraints on the spatial anisotropy and
the origin of HVSs.

\end{abstract}


\section{INTRODUCTION}

	Unbound HVSs were predicted by \citet{hills88} as the natural consequence of
the massive black hole (MBH) in the Galactic center. Following the discovery of the
first HVS \citep{brown05}, observers have reported the discovery of at least 16
unbound HVSs and evidence for a similar number of bound HVSs ejected by the same
mechanism \citep{hirsch05, edelmann05, brown06, brown06b, brown07a, brown07b,
brown08c}. Follow-up observations of 4 HVSs establish they are main sequence B stars
\citep{fuentes06, bonanos08, przybilla08, przybilla08b, lopezmorales08} like the
S-stars orbiting Sgr A$^*$ today \citep{ghez03, eisenhauer05, martins08}.
	Although not all unbound stars are necessarily HVSs -- the massive B star HD
271791 is the first example of an unbound ``hyper-runaway'' ejected from the outer
disk \citep{heber08, przybilla08c} -- runaway ejection velocities are limited to
$\sim$300~km~s$^{-1}$ for 3 M$_{\sun}$ stars \citep{leonard88, leonard90, leonard91,
leonard93, portegies00, davies02, gualandris05}.  Thus the 14 unbound 2.5-4
M$_{\sun}$ stars found in the \citet{brown07b, brown08c} targeted surveys are almost
certainly HVSs ejected from the Galactic center.

	Remarkably, 8 of the 14 HVSs in the \citet{brown07b, brown08c} targeted
surveys are located in just two constellations, Leo and Sextans, even though the
surveys cover $1/5^{\rm th}$ of the sky.  This spatial anisotropy is almost
certainly linked to the origin of the HVSs.

	In \S 2 we show that the observed distribution of HVSs on the sky is
anisotropic at the 3-$\sigma$ level.  In \S 3 we discuss plausible explanations for
the observed anisotropy of HVSs.

\section{OBSERVED ANISOTROPY}

\subsection{Sample}

	We consider the 14 HVSs from the combined surveys of \citet{brown07b,
brown08c}.  Our surveys use the MMT telescope to measure radial velocities for stars
with the colors of 2.5-4 M$_{\sun}$ stars. Heliocentric velocities are converted to
Galactocentric velocities assuming that the local rotation speed is 220 km s$^{-1}$
and that the Sun moves at (U,V,W)=(10, 5.2, 7.2) km s$^{-1}$ relative to the local
standard of rest \citep{dehnen98}.  The original HVS survey \citep{brown07b} is
100\% complete for stars with $17<g'_0<19.5$ over 7300~deg$^2$ covered by the Sloan
Digital Sky Survey Data Release 6.  The new HVS survey \citep{brown08c} is 59\%
complete for stars with $19.5<g'_0<20.5$ over the same region of sky.

\begin{figure}			% FIGURE 1: POLAR SKY MAP
 \plotone{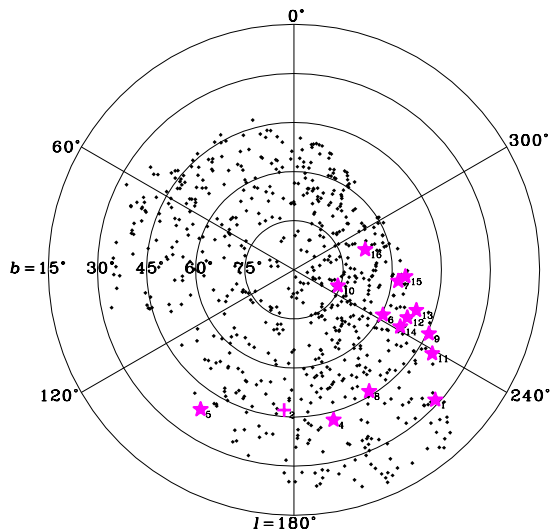}
 \caption{ \label{fig:polar}
	Polar projection, in Galactic coordinates, showing the 14 unbound HVSs ({\it
stars}) and the 693 other stars ({\it diamonds}) in our HVS survey \citep{brown08c}
covering the north Galactic cap.  HVS2, while not part of the survey, is also marked
({\it plus sign}).}
 \end{figure}

	Figure \ref{fig:polar} plots the spatial distribution of stars with observed
velocities.  The combined HVS survey contains 693 stars and 14 HVSs in the 7300
deg$^2$ Sloan region covering the north Galactic cap. HVS2 \citep{hirsch05} is also
located in this region (see Fig.\ \ref{fig:polar}), however it falls outside our
color / magnitude criteria. Thus we exclude HVS2 from this analysis.

\subsection{Significant Anisotropy}

\begin{figure}			% FIGURE 2: GLON GLAT DISTR
 \plotone{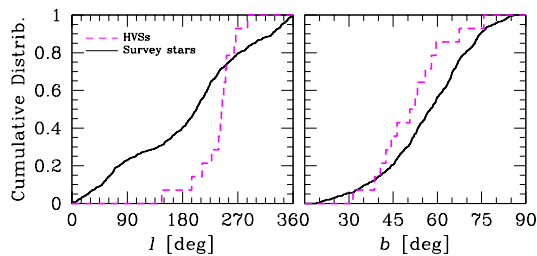}
 \caption{ \label{fig:gll}
	Cumulative distributions of Galactic $l$ and $b$ of the 14 HVSs ({\it
dashed lines}) and the 693 other stars ({\it solid lines}) in our HVS survey
\citep{brown08c}.}
 \end{figure}

	Figure \ref{fig:gll} plots the cumulative Galactic longitude and latitude
distributions of the HVSs and the other survey stars. Kolmogorov-Smirnov (K-S) tests
find 0.007 and 0.11 likelihoods that the HVSs are drawn from the same longitude and
latitude distributions, respectively, as the survey stars.  Thus the distribution of
HVS longitudes appears anisotropic at the 3-$\sigma$ level.

\begin{figure}			% FIGURE 3: SIGNIFICANCE
 \plotone{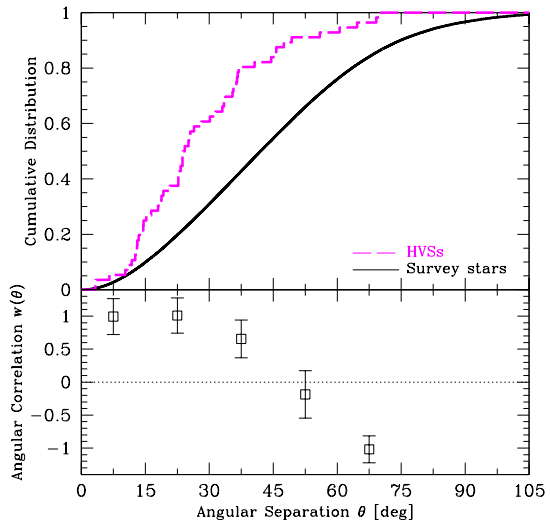}
 \caption{ \label{fig:angcor}
	Upper panel: cumulative distribution of angular separations, $\theta$.
Lower panel: two-point angular correlation function $w(\theta)$ of the HVSs with
respect to sets randomly drawn from the survey region. }
 \end{figure}

	As a second test, we explore the anisotropy in terms of the distribution of
angular separations, $\theta$, of the HVSs compared to the survey stars. Because the
new HVS survey is not yet complete, we calculate $\theta$'s for all unique pairs of
stars in the new and original surveys separately.  The original survey includes
HVS4-HVS10; a K-S test finds a 0.031 likelihood that those HVSs are drawn from the
same distribution of $\theta$ as the original survey stars.  The new survey includes
HVS1 and HVS11-HVS16; a K-S test finds a $7\times10^{-9}$ likelihood that those HVSs
are drawn from the same distribution of $\theta$ as the new survey stars.  Figure
\ref{fig:angcor} plots the $\theta$'s of both surveys concatenated together.  The
likelihood of the combined set of HVSs is $7\times10^{-8}$; thus the distribution of
HVS angular separations differs from the distribution of survey star angular
separations at the 5-$\sigma$ level.

	As a third test, we measure the clustering of HVSs using the two-point
angular correlation function $w(\theta)$.  We use a Monte Carlo estimator
\citep{landy93} and compare the observed HVSs against 10$^5$ sets randomly drawn
from the survey region.  The lower panel of Figure \ref{fig:angcor} plots the
resulting $w(\theta)$ in 15$^{\circ}$ bins.  Errorbars are determined by Poisson
statistics.  HVSs are clustered at small angular separations $\theta<45^{\circ}$ and
missing at large angular separations $\theta>60^{\circ}$ with $\sim$3.5-$\sigma$
significance.

\subsection{Velocity Dependence}

	We now consider the spatial anisotropy of lower velocity stars that may also
be HVSs.  \citet{brown08c} identify 4 ``possible HVSs,'' stars that are bound in the
\citet{kenyon08} potential model but unbound in the \citet{xue08} potential model.  
Adding the 4 possible HVSs to the above analysis reduces the significance of the
anisotropy to the 2-$\sigma$ level. There are also 8 possibly ``bound HVSs,'' stars
with $v_{rf}>+275$ km s$^{-1}$ that are significant outliers from the overall
velocity distribution \citep{brown08c}.  Adding the bound HVSs to the above analysis
yields an insignificant anisotropy.  Thus lower velocity stars have a more isotropic
distribution, a trend noted previously in \citet{brown07a}.

\subsection{HVS Pairs}

	There are 3 pairs of unbound HVSs with angular separations less than
3.5$^{\circ}$ (see Fig.\ \ref{fig:polar}): HVS7 \& HVS15 near
($l,b$)=(265$^{\circ}$, 55$^{\circ}$), HVS12 \& HVS13 and HVS12 \& HVS14 near
($l,b$)= (245$^{\circ}$, 52$^{\circ}$).
	Any physical association between the individual HVSs, however, appears
unlikely.  HVS7 \& HVS15 are separated by $2.5^{\circ}$ but have velocities and
distances that imply a $\sim$70 Myr difference in travel time from the Galactic
Center \citep[see Fig.\ 3 of][]{brown08c}.  HVS12 has a 429 km s$^{-1}$ minimum
rest-frame velocity very similar to that of HVS13 and HVS14, but it has half the
distance of the other two HVSs. Thus none of the HVS pairs share a common ejection
event.

\section{ORIGIN OF THE SPATIAL ANISOTROPY}

	We observe that the spatial anisotropy of unbound HVSs is statistically 
robust, that lower velocity HVSs are systematically more isotropic, and that 
apparent close pairs of HVSs are physically unrelated.  Possible explanations for 
the observations include:

\vskip 2pt

\noindent {\it Selection Effect.} Previously, we argued that the HVS anisotropy may
be a selection effect of our magnitude-limited survey and the Sun's off-center
location in the Galaxy \citep{brown07a}.  However, this selection effect can account
only for an extra $\sim$10\% HVSs in the anti-center hemisphere, not all of the HVSs
in the anti-center hemisphere. Moreover, the observed HVSs cluster around
$l=240^{\circ}$, not $l=180^{\circ}$.

\vskip 2pt

\noindent {\it Runaways.} Runaway stars like HD 271791 may contaminate the
population of HVSs.  However, we expect runaways ejected from the disk to have an
isotropic distribution in Galactic longitude, as demonstrated by the
\citet{martin06} Hipparcos-selected sample of runaway B stars.  Moreover, because
runaways are systematically ejected at low velocities \citep[e.g.,][]{portegies00},
the fastest runaways are those ejected in the direction of Galactic rotation and
thus preferentially found at low Galactic latitudes.  Thus the expected distribution
of runaway longitudes and latitudes are contrary to the observed distribution of
HVSs.

\vskip 2pt

\noindent {\it Large Scale Structure.} The distribution of Local Group dwarf
galaxies is anisotropic, possibly due to a tidal origin \citep[e.g.,][]{metz08}.  A
tidal debris origin appears supported by the clumping of HVS travel times around
100-200 Myr, however the travel times are simply a product of the HVS's $\sim$500 km
s$^{-1}$ velocities and our magnitude-limited survey depth of 50-100 kpc.  HVS
travel times are in fact problematic for a tidal debris origin because the times are
a significant fraction of the stars' main sequence lifetimes, and multiple
(gas-rich) tidal disruption events would be required to explain the full
$2\times10^8$ yr span of HVS travel times.  No dwarf galaxy in the Local Group
travels with radial velocities comparable to the unbound HVSs; known dwarf galaxy
remnants like the Sgr stream \citep{ibata94} are bound.  We thus consider tidal
debris an unlikely explanation for the observed set of HVSs \citep[however,
see][]{abadi08}.

\vskip 2pt

\noindent {\it Binary Black Hole.} While an equal-mass binary MBH is ruled out in
the Galactic Center \citep{reid04}, theorists speculate that the massive star
clusters in the Galactic Center form intermediate mass black holes (IMBHs) in their
cores.  If such IMBHs exist, dynamical friction causes them to in-spiral into the
central MBH, preferentially ejecting HVSs from their orbital planes.  Thus the
expected signature of a IMBH in-spiral is a ring of HVSs around the sky
\citep{gualandris05, levin06, sesana06}.  \citet{baumgardt06} argue, however, that
stellar interactions perturb the orbital plane of an in-spiraling IMBH; the
resulting HVS distribution in this scenario may in fact be isotropic.  Moreover, a
single IMBH in-spiral event happens on timescales 10-100$\times$ shorter than the
observed span of HVS travel times; multiple IMBH in-spiral events are required to
explain the observed HVSs.

\vskip 2pt

\noindent {\it Galactic Center Structure.} The Galactic center contains many
well-defined structures.  As illustrated in \citet{paumard06}, the molecular gas
circum-nuclear disk and the ionized northern arm are roughly aligned with the plane
of the Milky Way.  The gaseous minispiral is perpendicular to the plane of the Milky
Way.  Notably, the stellar disk 0.1 pc from the MBH is roughly perpendicular to the
gaseous components \citep{lu08}.
	The stellar disk contains massive stars \citep{tanner06, paumard06},
possibly formed in-situ from a gas accretion disk \citep{genzel03, levin03}.  
Dynamical interactions between a pair of stellar disks may scatter stars in towards
the MBH, explaining both the S-stars and the HVSs \citep{lockmann08b, perets08c}.  
Clearly, the Galactic center contains non-isotropic distributions of stars and gas
which may provide a natural source for the observed anisotropy of HVSs ejected from
the Galactic center.  However, it is unclear if the observed structures can persist
long enough to explain the anisotropic distribution of HVSs.

%\noindent {\it Galactic Potential.} The Galactic potential maps the initial HVS
%ejection velocities from the Galactic center to the final velocities observed in 
%the outer halo.  HVSs ejected at the highest velocities escape the Galaxy while 
%lower velocity ejections remain bound.  Because many HVSs are marginally unbound, 
%an anisotropic potential can lead to an observed anisotropic spatial distribution 
%of HVSs.  NOW IN SEPARATE PAPER PER THE REFEREE.

\section{CONCLUSION}

	Unbound HVSs are spatially anisotropic at the 3-$\sigma$ level.  The 
anisotropy is most significant in Galactic longitude, and not in latitude.
Lower velocity HVSs are systematically more isotropic, and apparent close
pairs of HVSs are physically unrelated.  

	The observed distribution of HVSs is linked to the origin of the HVSs.  
\citet{abadi08} propose a tidal debris explanation, although this appears difficult
to reconcile with all the observations.  We investigate other physical models for
the anisotropy in a separate paper.  In the future, measuring the distribution of
bound and unbound HVSs over the southern sky will allow us to better constrain the
anisotropy and the origin of HVSs.

\acknowledgements

	This work is based on observations obtained at the MMT Observatory, a joint
facility of the Smithsonian Institution and the University of Arizona.  This
research makes use of NASA's Astrophysics Data System Bibliographic Services.  We
thank the anonymous referee and Oleg Gnedin for helpful comments.  This work was
supported by the Smithsonian Institution.

{\it Facilities:} {MMT (Blue Channel Spectrograph)}

	% REFERENCES
%\clearpage
%\bibliographystyle{/home/wbrown/lib/apj} \bibliography{/home/wbrown/text/RefHS} 

	% FIGURES
%\clearpage

\end{document}